\documentclass[aps,twocolumn,pra,showpacs,superscriptaddress]{revtex4-1}

\usepackage{amsfonts,amsthm}
\usepackage{subfigure}
\usepackage{amssymb}
\usepackage{amscd}
\usepackage{amsmath}    % need for subequations
\usepackage{multirow}
\usepackage{graphicx}% Include figure files
\usepackage{dcolumn}% Align table columns on decimal point
\usepackage{bm}% bold math
\usepackage{cleveref}
\newtheorem{definition}{Definition}

\newtheorem{theorem}{Theorem}

\input{Qcircuit}

\begin{document}
\title{Counterfactual universal quantum computation}

\author{Zhu Cao}
\email{caozhu@ecust.edu.cn}
\address{Key Laboratory of Advanced Control and Optimization for Chemical Processes of Ministry of Education, East China University of Science and Technology, Shanghai 200237, China}
\address{Shanghai Institute of Intelligent Science and Technology, Tongji University, Shanghai 200092, China}

\begin{abstract}
Universal quantum computation is usually associated with interaction among two-level quantum subsystems, as this interaction is commonly viewed as a necessity to achieve universal quantum computation. In this work, we show that, contrary to this intuition, universal quantum computation can be achieved without interaction among initially independent two-level quantum subsystems. We call it counterfactual universal quantum computation. As special cases, we show how to achieve counterfactual communication of quantum states, counterfactual quantum swapping, and counterfactual quantum erasure codes. To ease practical implementation, we analyze counterfactual universal quantum computation with realistic devices, including the effects of finite execution time, photon loss, and atom missing. Besides the theoretical interest of illustrating the mysterious and counterintuitive nature of quantum physics, our work has practical applications to color imaging of ancient arts, upon which light is forbidden to shine.  
\end{abstract}

\maketitle

\section{Introduction}

The term ``counterfactual'' was first coined to describe the phenomenon of interaction-free measurement, where an object is determined to be present or absent without interaction with any probing particle. The first interaction-free measurement scheme \cite{dicke1981interaction,elitzur1993quantum} had a limited efficiency of 50\% and was subsequently improved to 100\% efficiency \cite{kwiat1995interaction} by the quantum Zeno effect \cite{misra1977zeno,peres1980zeno,agarwal1994all}, an effect where a particle is frequently weakly measured so that the state of the particle stays unchanged with a high probability. The scheme was later extended to the scenario of quantum cryptography \cite{noh2009counterfactual,ren2011experimental,brida2012experimental,liu2012experimental},  quantum interrogation \cite{kwiat1999high}, and entanglement distribution \cite{elitzur2015quantum,aharonov2018interaction}. 

In this work we develop a counterfactual universal quantum computation (CUQC) scheme that has the following counterfactual property: it can accomplish universal quantum computation with no interaction among initially independent two-level quantum subsystems throughout the computation process. This is counterintuitive as it seems impossible for such a scheme to realize entanglement among two-level quantum subsystems, which is a necessary condition for universal quantum computation. In addition, so far, there has been no scheme that can fulfill this counterfactual property.
Note that the term ``counterfactual quantum computation'' has had a different meaning \cite{hosten2006counterfactual,kong2015experimental}, which is that the quantum computation itself is in a superposition of ``run'' and ``not run,'' although the two-level quantum subsystems have notable interactions. The correct outcome of the quantum computation can be obtained with high probability even when the quantum computation runs with negligible probability. The basic ingredient is also the interaction-free measurement.

The  CUQC scheme was motivated by the recent development of counterfactual communication protocols, which starts with the Salih {\it et al.} protocol for counterfactual communication of a classical bit \cite{salih2013protocol}, which was later experimentally demonstrated \cite{cao2017direct}. The counterfactuality of this pioneering scheme is however debatable \cite{vaidman2014comment,salih2014salih,salih2018laws}. The main opposing argument is that, according to a stricter counterfactuality definition  \cite{vaidman2013past}, the scheme fails to be counterfactual. Recently, Aharonov and Vaidman \cite{aharonov2019modification} provided another scheme for the same task which is counterfactual even under the stricter counterfactuality definition  \cite{vaidman2013past}. Their scheme can be seen as a modification of the Salih {\it et al.} scheme, with the difference that an additional double-sided mirror is added to the setup.

As special cases of the CUQC scheme, we can achieve counterfactual communication of a quantum state, counterfactual quantum swapping, and counterfactual quantum erasure codes. In counterfactual communication of a quantum state, a quantum state is transmitted counterfactually from one party to another party. Note that a debatable scheme for counterfactual communication of a quantum state was previously proposed \cite{li2015direct,vaidman2016comment}. The opposing argument is also that the scheme is not counterfactual under the stricter counterfactuality definition  \cite{vaidman2013past}.
Our result is counterfactual even under the stricter counterfactuality definition \cite{vaidman2013past}, and can be seen as a quantum generalization of Aharonov and Vaidman's scheme \cite{aharonov2019modification}. In counterfactual quantum swapping, two quantum states are counterfactually swapped. This special case recovers a recent result of Li {\it et al.} \cite{li2019counterfactual}. In counterfactual quantum erasure code, a highly entangled erasure code is prepared from separable states counterfactually. The code ensures that even if some of the qubits are erased, the correct logical quantum state can still be recovered through error correction.

For experimental implementations, we consider practical aspects such as finite size and loss. In particular, we
find that the fidelity of the computation remains high under finite execution time, but the efficiency of the computation
has a notable change under finite execution time. We also find that the fidelity and the efficiency are both sensitive to a small probability of photon loss but  insensitive to a small probability of atom missing. Here photons and atoms are media that are utilized in our CUQC scheme. We also discuss the dependence of the efficiency and the fidelity on the size of the quantum computation circuit. In particular, we show in detail how the fidelity deteriorates with an increase of the quantum circuit size if no error correction is performed. From an orthogonal perspective, we analyze the performance of the CUQC scheme between using a single atom and multiple atoms, and find that multiple atoms can significantly speed up the computation process in certain cases. 

The road map for the rest of the paper is as follows. In Sec.~\ref{sec:preliminaries} we review some definitions and theorems that will be needed later in the paper. In Sec.~\ref{sec:scheme} we present the CUQC scheme. In Sec.~\ref{sec:examples} we give three examples of the CUQC scheme. In Sec.~\ref{sec:extensions} we consider practical aspects of the CUQC scheme. In Sec.~\ref{sec:conclusion} we provide a summary and show some promising future research directions.

\section{Preliminaries}
\label{sec:preliminaries}
The following is the definition of universal quantum computation.
\begin{definition}[Universal Quantum Computation \cite{deutsch1989quantum}]
A universal quantum computation is an arbitrary unitary transformation on a discrete Hilbert space spanned by the set of all quantum states of a collection of qubits.
\end{definition}
We will use the following important property of universal quantum computation.
\begin{theorem}[One-qubit gates and CNOT gates are universal \cite{divincenzo1995two}]
\label{thm:universal}
A universal quantum computation can be realized by a set of one-qubit gates and CNOT gates.
\end{theorem}

Next is the definition of the presence of a quantum particle.
\begin{definition}[Presence for a quantum particle \cite{vaidman2013past}]
A quantum particle is said to be present at a point $P$ if the weak value of the particle at $P$, measured by a weak measurement, is nonzero. Otherwise, the particle is said to be absent at $P$.
\label{def:presence}
\end{definition}
A simple method, called two-state vector formalism, can be used to test whether the weak value is nonzero. It can be stated as follows.
\begin{theorem}[Two-state vector formalism \cite{vaidman2013past}]
\label{thm:twovec}
The weak value of a quantum particle is nonzero at a point $P$ if and only if both the forward- and backward-evolving wave functions of the quantum particle do not vanish at $P$.
\end{theorem}

\begin{figure}[htb]
\centering \includegraphics[width=8.5cm]{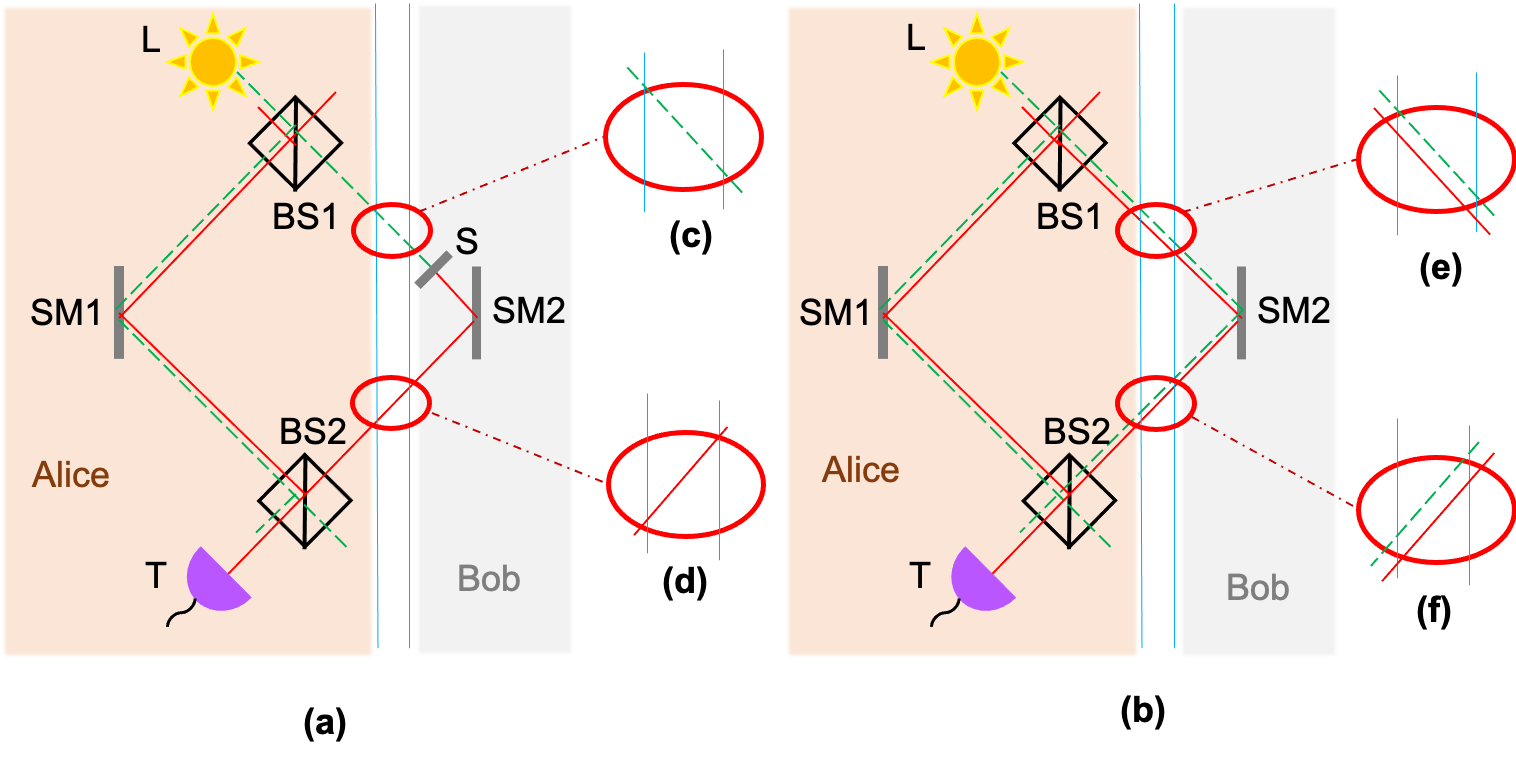}
\caption{Illustration of the two-state vector formalism. Consider a photon that is emitted from a light source L and received by a detector T. The forward-evolving (backward-evolving) wave function of this photon is depicted by green dashed  (red solid) lines. The region between two parallel blue lines is the transmission channel. Alice (Bob) is on the left (right) of the channel. Two cases are considered: (a) Bob has a shutter S and (b) Bob has no shutter. (c) and (d) The forward- and backward-evolving wave functions do not overlap. (e) and (f)  The forward- and backward-evolving wave functions overlap.
Here L denotes the light source, BS1 and BS2 are beam splitters, SM1 and SM2 are single-sided mirrors, T is a single-photon detector, and S is a shutter. } 
\label{fig:interferometer}
\end{figure}

As an example to illustrate this method and the concepts of forward- and backward-evolving wave functions, consider the quantum optical setup in Fig.~\ref{fig:interferometer}(a), which is essentially an interferometer. The transmission channel is depicted as the region between two parallel blue lines. Alice is on the left side of the channel and Bob is on the right side of the channel. 
Given a source and a detector, the forward- and backward-evolving wave functions can be defined as follows.
The forward-evolving wave function is defined as the wave function that starts with the source and is depicted by green dashed lines. In this setup, it starts with the light source L and splits into two paths at the first beam splitter BS1.  The right path hits Bob's shutter S. The left path is reflected by the single-sided mirror SM1 and then splits into two paths at BS2. The backward-evolving wave function is defined as the wave function that starts with the detector and is depicted by red solid lines. In the current setup, it starts with the detector T and splits into two paths at BS2. The right path is first reflected by the single-sided mirror SM2 and then hits Bob's shutter S. The left path is first reflected by the single-sided mirror SM1 and then splits into two paths at BS1. 
It can be seen that the forward- and backward-evolving wave functions do not overlap in the transmission channel, as shown in Figs.~\ref{fig:interferometer}(c) and~\ref{fig:interferometer}(d), indicating there is no photon in the transmission channel. 

Next we examine the opposite situation, as shown in Fig.~\ref{fig:interferometer}(b). The optical setup is almost the same as the previous setup, except that Bob's shutter S is removed. The forward-evolving wave function still starts at L and splits into two paths at BS1. The left path is the same as the previous situation. The right path is reflected by SM2 and splits into two paths at BS2. The backward-evolving wave function still starts at the detector T and splits into two paths at BS2. The left path remains the same as the previous situation. The right  path is reflected by SM2 and then splits at BS1 into two paths. It can be seen that the forward- and backward-evolving wave functions overlap in the transmission channel, as shown in Figs.~\ref{fig:interferometer}(e) and \ref{fig:interferometer}(f); hence the photon is present in the channel.

\section{Counterfactual universal quantum computation scheme}
\label{sec:scheme}
In this section we present our CUQC scheme. We begin with the definition of counterfactuality for universal quantum computation in Sec.~\ref{sec:defCUQC}. Then we build the CUQC scheme in two steps. In the first step, we design a counterfactual special CNOT gate. The exact definition of ``special'' is given in Sec.~\ref{sec:CNOT}. For now, we content ourselves by viewing a special CNOT gate as a weakened version of a CNOT gate. In the second step, given in Sec.~\ref{sec:SingleCompute}, we show how to achieve the CUQC scheme through the combination of single-qubit operations and counterfactual special CNOT gates.

\subsection{Definition of CUQC}
\label{sec:defCUQC}
A CUQC scheme has the following three properties.
\begin{enumerate}
\item It can achieve universal quantum computation.
\item Initially, the two-level quantum subsystems are independent.
\item There is no interaction among the two-level quantum subsystems during the quantum computation.
\end{enumerate}

\begin{figure}[htb]
\centering \includegraphics[width=4.5cm]{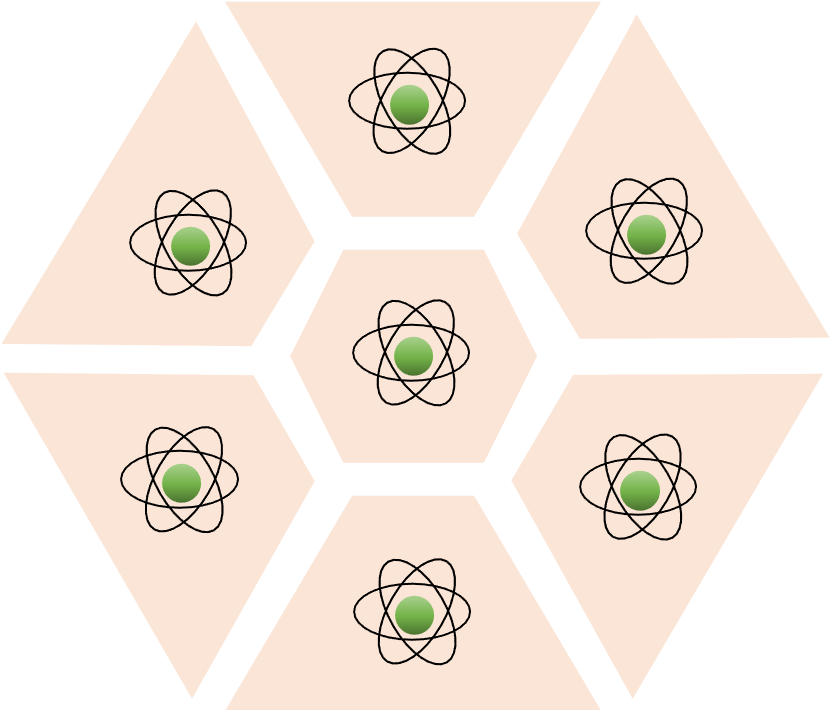}
\caption{Illustration of CUQC where there is no interaction among the two-level quantum subsystems. Each two-level quantum subsystem is confined in an isolated chamber (pink polygon). In addition, no physical particles are present at any point out of the chamber (white region) at any time during the computation.}
\label{fig:definition}
\end{figure}

The first two conditions are quite straightforward, but the third condition has a few subtleties. As depicted in Fig.~\ref{fig:definition}, \emph{no interaction} in this context means that every two-level quantum subsystem is restricted in its isolated chamber and away from other two-level quantum subsystems. In addition, there is no physical particle, such as a photon or a phonon, that can potentially carry information from one two-level quantum subsystem to another two-level quantum subsystem, present at any point outside the chambers during the computation. The presence of a physical particle is defined by Definition~\ref{def:presence}.

\subsection{Counterfactual special CNOT gate}
\label{sec:CNOT}
In this section we design a counterfactual special CNOT gate. 
This CNOT gate is special in that there is a constraint on which qubit can be the control qubit and which qubit can be the 
target qubit. More precisely, only qubits that are realized by atoms can be the control qubits and only qubits that are realized by
photons can be the target qubits. This CNOT gate is counterfactual in that there is no interaction between the photon and the atom during this gate operation.

Our counterfactual special CNOT gate is inspired by a recent scheme that counterfactually communicates a classical bit  \cite{aharonov2019modification}, which we now review.
Counterfactuality here means that no particle is present in the transmission channel during the communication under Definition \ref{def:presence}. This in particular implies that the probability that a photon is detected by a photon detector in the transmission channel at any time should be zero. The scheme consists of three ingredients. 

\begin{figure}[htb]
\centering \includegraphics[width=8.5cm]{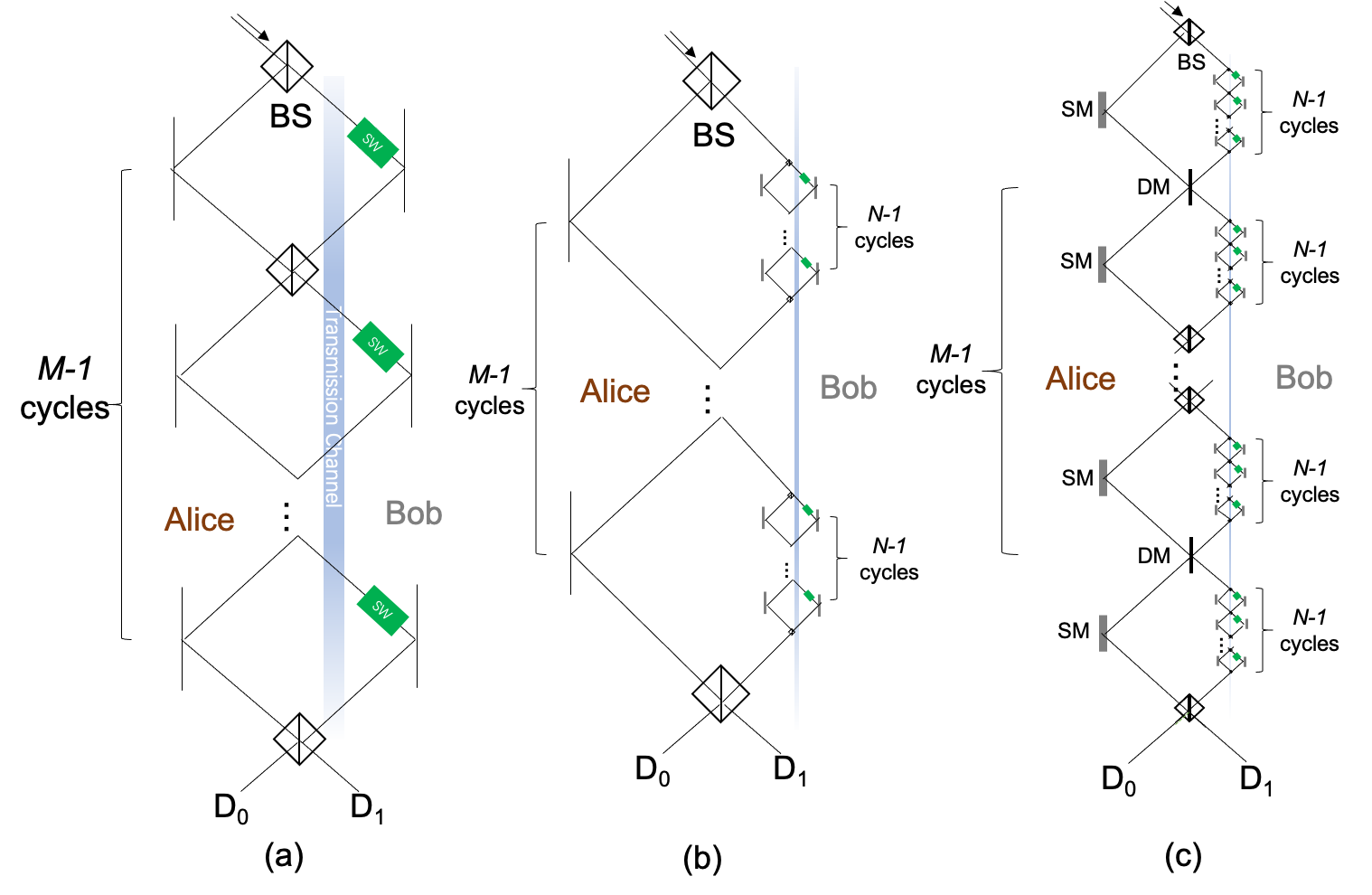}
\caption{(a) Quantum Zeno effect. For large $M$, if the upper arms are blocked (unblocked), the photon always goes to D$_0$ (D$_1$).  (b) Chained quantum Zeno effect. The photon has zero probability to be detected in the transmission channel for both the block and unblock cases. (c) A double-sided mirror is added to eliminate the presence of the photon in the transmission channel. Here BS denotes a beam splitter, SM is a single-sided mirror, DM is a double-sided mirror, D$_0$ and D$_1$ are single-photon detectors, and SW is Bob's switch.} 
\label{fig:scheme}
\end{figure}

The first ingredient is the quantum Zeno effect. It is realized by a tandem interferometer with $M$ beam splitters (BSs) of reflectivity $\cos^2(\pi/2M)$, as shown in Fig.~\ref{fig:scheme}(a). 
The transmission channel cuts the interferometer into two halves. Alice is on the left side and Bob is on the right side. Bob aims to communicate a classical bit to Alice.
All beam splitters are on Alice's side. Bob has a switch SW to determine whether the upper arms of the cycles in the tandem interferometer are blocked. Alice's two detectors D$_0$ and D$_1$ are placed after the last BS.
Initially, a photon comes from the upper left side of the first BS, as the arrow in Fig.~\ref{fig:scheme}(a) shows.
 If the upper arms of the tandem interferometer are blocked, for each cycle only the reflected component [with proportion $\cos^2(\pi/2M)$ according to the property of the BS] survives to the next cycle, and hence the photon goes to detector D$_0$ with probability 
\begin{equation}
\textsf{Prob}(\textrm{D}_0)=(\cos^2(\frac{\pi}{2M}))^M \approx (1 -   (\frac{\pi}{2M})^2/2  )^{2M} \approx e^{ - \pi^2/4M}.
\end{equation}
 When $M$ goes to infinity, this probability goes to 1. Consequently, D$_0$ always clicks while D$_1$ never clicks. If the upper arms are not blocked, 
the probability that the photon goes across the transmission channel is $\sin^2(i\pi / 2M)$  for the $i$th cycle, and hence
 the photon goes to detector D$_1$ with probability 
 \begin{equation}
\textsf{Prob}(\textrm{D}_1)= \sin^2(\frac{M\pi}{2M} ) =1.
 \end{equation}
 Hence, the block status of the upper arms (block or pass), a classical bit of Bob, is transmitted to the click status of the detectors (D$_0$ clicks or D$_1$ clicks), a classical bit of Alice. If this communication is also counterfactual, then Bob's goal is achieved. However, when the upper arms are not blocked, it can be shown that for the last cycle, the probability of the photon to be detected in the transmission channel is $\sin^2[ (M-1)\pi / 2M]\approx 1$ for large $M$. Hence, the communication is unfortunately not counterfactual.
 
\begin{figure}[htb]
\centering \includegraphics[width=8.5cm]{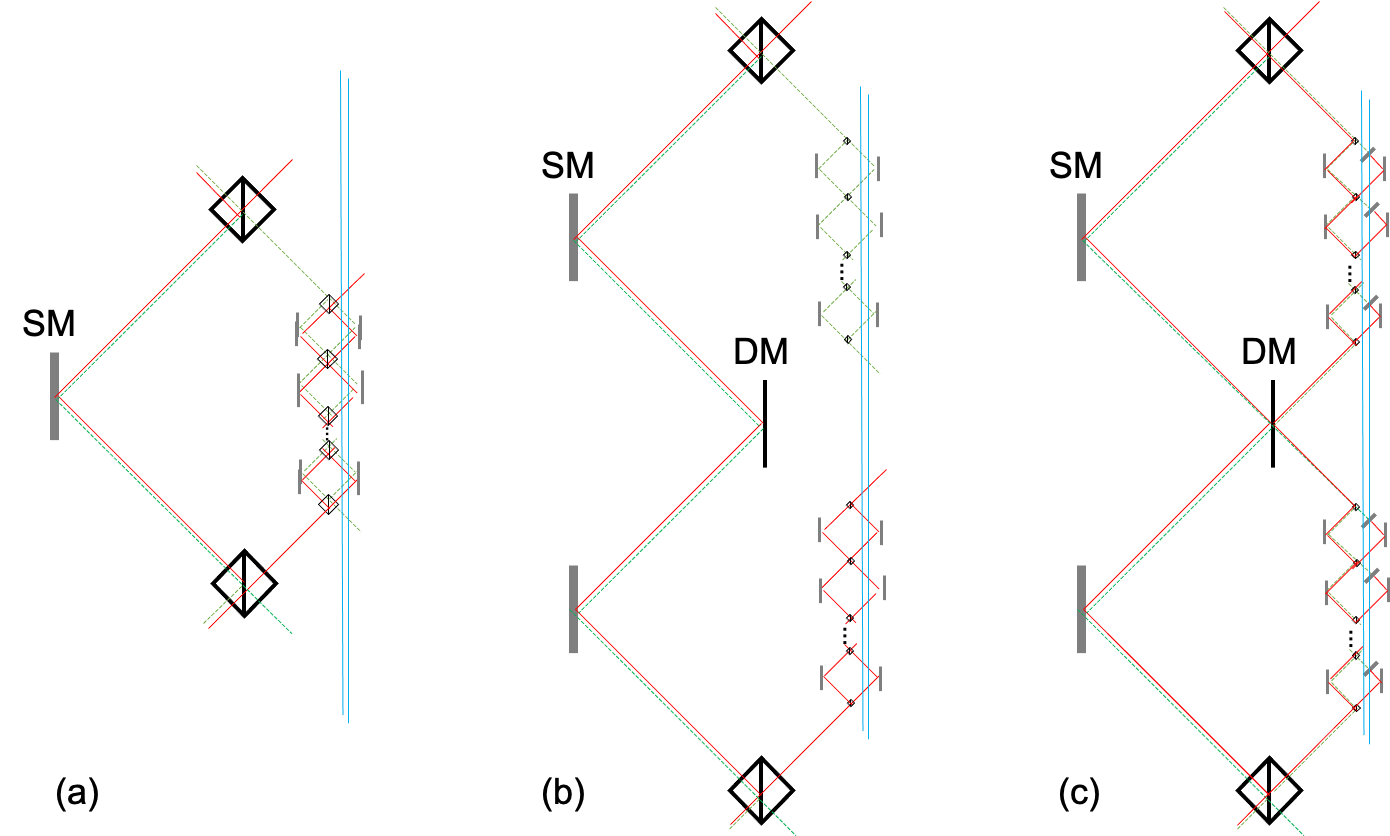}
\caption{Forward- and backward-evolving wave functions of one outer cycle (a) without Bob's shutter or a double mirror, (b) without Bob's shutter but with a double mirror, and (c) with both Bob's shutter and a double mirror.  The forward-evolving wave function is represented by green dashed lines and the backward-evolving wave function is shown by red solid lines. Here SM denotes a single-sided mirror and DM a double-sided mirror.} 
\label{fig:twovector}
\end{figure}

To avoid a nonzero probability of photon detection in the transmission channel, the second ingredient, the chained quantum Zeno effect, is introduced. It is realized by nesting a tandem interferometer inside each cycle of the outer tandem interferometer (\emph{outer cycle} for short), as shown in Fig.~\ref{fig:scheme}(b). Each nested inner tandem interferometer consists of $N$ beam splitters with reflectivity $\cos^2(\pi/2N)$. Again, we assume $N$ goes to infinity. If the upper arms of all cycles of an inner tandem interferometer (\emph{inner cycles} for short) are blocked (unblocked), the photon goes to the left (right) of the last BS of the inner tandem interferometer, which is equivalent to a pass (block) status of an outer cycle, and hence D$_1$ (D$_0$) will click. The overall effect is that the block status of the inner cycles (block or pass) corresponds to the click status of the detectors (D$_1$ clicks or D$_0$ clicks) and therefore a classical bit is communicated from Bob to Alice. For the block case, the probability of photon detection in the transmission channel is bounded by $\sin^2(\pi/ 2N)$, and for the non-block case, it is bounded by $\sin^2(\pi/ 2M)$. Therefore, the probabilities of the photon passing through the channel for both the block case and the nonblock case go to zero as $M$ and $N$ go to infinity. However, for the nonblock case, the photon is still present in the transmission channel and hence counterfactuality is not achieved. This can be seen through Fig.~\ref{fig:twovector}(a), which depicts one outer cycle, in which the forward- and backward-evolving wave functions overlap in the transmission channel, indicating that the photon is present.

The critical idea of Ref.~\cite{aharonov2019modification} is 
the third ingredient, which joins two old outer cycles by a double-sided mirror to form a new outer cycle, as shown in Fig.~\ref{fig:scheme}(c). A new outer cycle now contains $2(N-1)$ inner cycles and there are  in total $M-1$ new outer cycles.
Note that this new setup still maintains the property that the block status of the inner cycles (block or pass) corresponds to the click status of the detectors (D$_1$ clicks or D$_0$ clicks). It remains to be
 shown that counterfactuality holds for both the block case and the pass case. The forward- and backward-evolving wave functions of one outer cycle are shown in Figs.~\ref{fig:twovector}(b) and (c). 
In Fig.~\ref{fig:twovector}(b), the upper arms are not blocked, while in Fig.~\ref{fig:twovector}(c), the upper arms are blocked. It can be seen that for both cases, the forward- and backward-evolving wave functions do not overlap in the transmission channel, so the photon is not present by Theorem \ref{thm:twovec}. Hence, this scheme achieves counterfactual communication of a classical bit.

\begin{figure}[htb]
\centering \includegraphics[width=8.5cm]{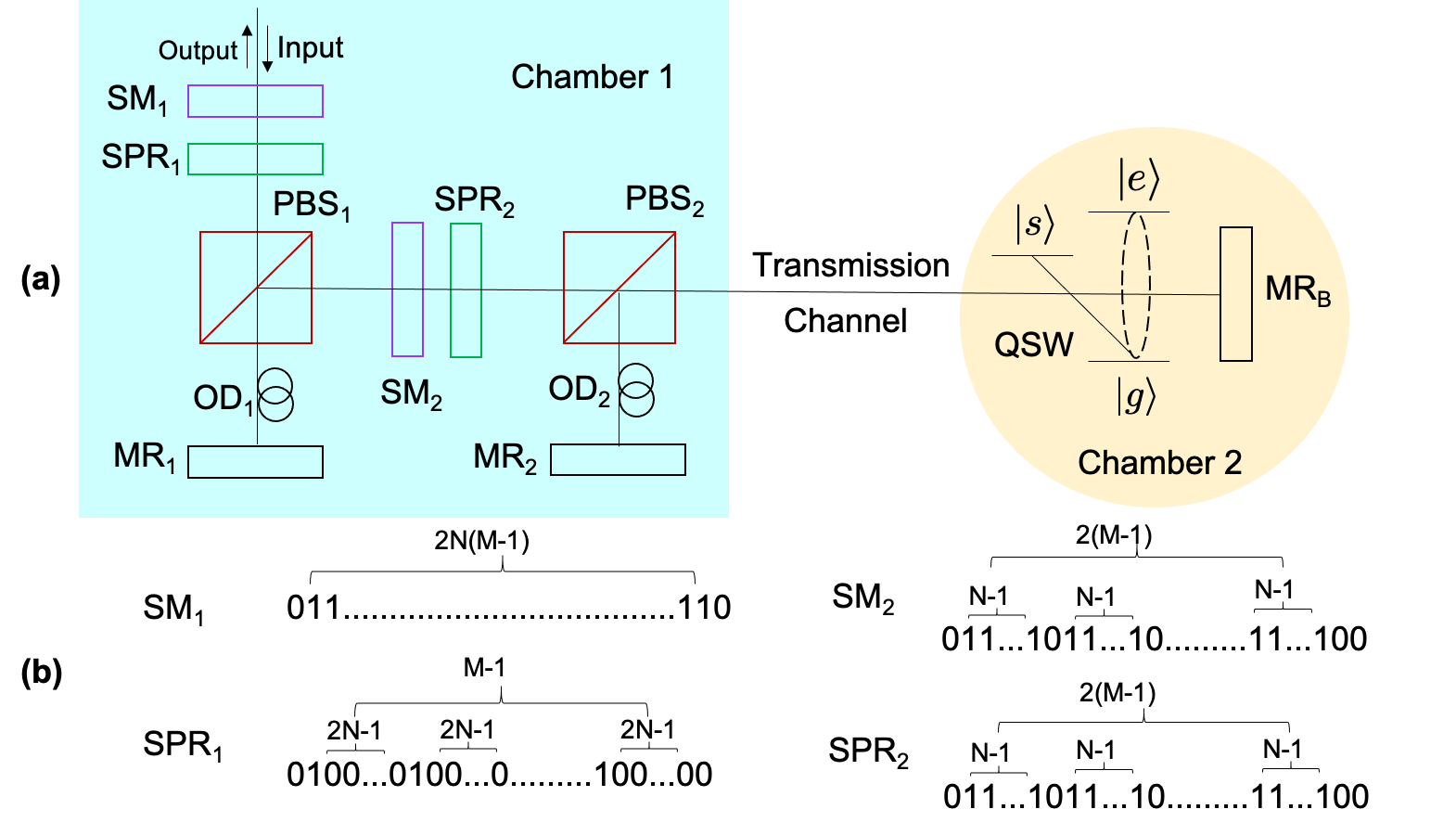}
\caption{(a) Optical setup of a counterfactual special CNOT gate (the detailed protocol is explained in the text): D, single-photon detector; SM, switchable mirror; SPR, switchable polarization rotator; MR, mirror; OD, optical delay; PBS, polarizing beam splitter; QSW, quantum switch; and $\ket{g}$, $\ket{e}$, and $\ket{s}$, quantum states of an atom qubit. (b) On-off states of switchable components. Here 0 denotes the off state and 1 denotes the on state.} 
\label{fig:setup}
\end{figure}

Our counterfactual special CNOT gate modifies the above scheme in two aspects. The first modification replaces Bob's classical switch with a quantum switch.
The quantum switch can be realized by an atom that contains three energy levels $\ket{e}$, $\ket{g}$, and $\ket{s}$  \cite{li2015direct}, as shown in the right panel of Fig.~\ref{fig:setup}(a).
When the atom is in the ground state $\ket{g}$, it absorbs any passing photon, changes its state to the second excited state $\ket{s}$, and stays there. When the atom is in the first excited state $\ket{e}$, 
it is unable to absorb photons. Hence, the ground state $\ket{g}$ acts as the block status and the first excited state $\ket{e}$ acts as the pass status.
The second modification replaces the beam splitters by polarization beam splitters and uses the polarization encoding for the photons. The horizontal polarization $\ket{H}$ and the vertical polarization $\ket{V}$ stand for logical $\ket{0}$ and logical $\ket{1}$, respectively. 

The complete setup is shown in Fig.~\ref{fig:setup}(a). Initially, all switchable components are turned off. A switchable component becomes a normal component when it is in the on state and disappears when it is in the off state.
After an input photon with polarization $\ket{H}$ passes a switchable mirror SM$_1$, SM$_1$ is turned on for $M-1$ outer cycles. 

At the start of each outer cycle, a switchable polarization rotator SPR$_1$ is turned on, which rotates the polarization of the photon with an angle $\beta_1=\pi/2M$.
Here a rotation with an angle $\beta_1$ means that the quantum state of the photon evolves
as 
\begin{equation}
\left(    
 \begin{array}{c}  
    X\\ 
    Y\\
  \end{array}
\right)
\rightarrow
  \left(
  \begin{array}{cc}  
    \cos \beta_1 & \sin\beta_1\\ 
   -\sin\beta_1& \cos \beta_1\\
  \end{array}  
  \right)
\left(  
 \begin{array}{c}  
    X \\ 
    Y\\
  \end{array}
  \right)
  \equiv
  \left(    
 \begin{array}{c}  
    X'  \\ 
    Y' \\
  \end{array}
\right)
  ,
\end{equation}
where $(X, Y)^T$ denotes the quantum state $X\ket{H}+Y\ket{V}$.
Then SPR$_1$ is  turned off and remains off for the rest of this outer cycle. 
The photon then splits into two components by a polarization beam splitter PBS$_1$, of which the vertical component passes a switchable mirror SM$_2$ and the horizontal component is reflected by a mirror MR$_1$. An optical delay OD$_1$ is put on the horizontal component so that both components return to PBS$_1$ at the same time.
The switchable mirror SM$_2$ is turned on for $N-1$ inner cycles.
At the start of each inner cycle, a switchable polarization rotator SPR$_2$ is turned on, rotates the polarization of the photon by an angle of $\beta_2=\pi/2N$, and is then turned off.  The photon then splits into two components at a polarization beam splitter PBS$_2$, of which the horizontal component passes the transmission channel and the vertical component is reflected by a mirror MR$_2$. An optical delay OD$_2$ is put on the vertical component so that both components return to PBS$_2$ at the same time. After passing the transmission channel, the photon is either blocked by Bob's atom or reflected by a mirror MR$_B$. After $N-1$ inner cycles, SM$_2$ is turned off to allow the photon to exit the inner cycle. The switchable mirror SM$_1$ then reflects the photon, with SPR$_1$ remaining off this time so that no polarization rotation is applied to the photon, mimicking the function of a double mirror. After another $N-1$ inner cycles, one outer cycle is finished. 

\begin{figure}[htb]
\centering \includegraphics[width=4cm]{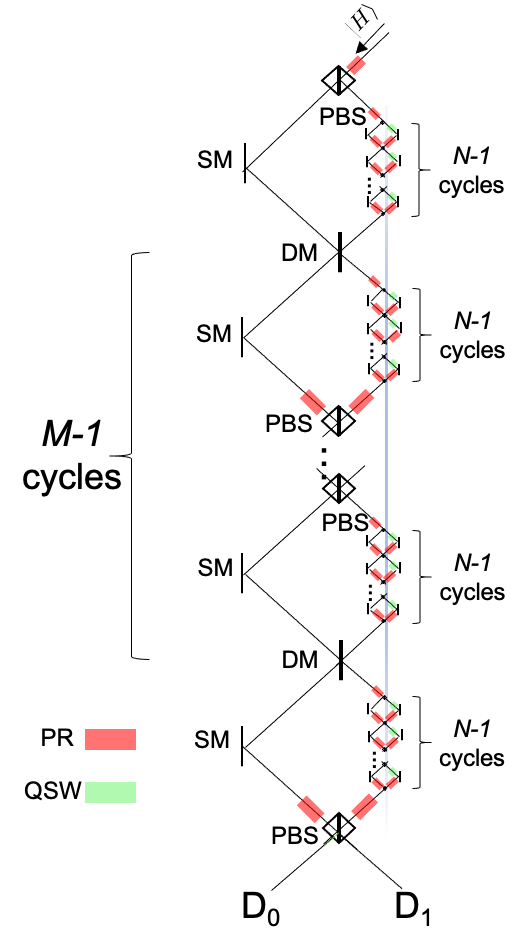}
\caption{Equivalent optical setup of a counterfactual special CNOT gate: SM, single-sided mirror; DM, double-sided mirror; PBS, polarization beam splitter; PR, polarization rotator; and QSW, quantum switch. } 
\label{fig:convert}
\end{figure}

After $M-1$ such outer cycles, SM$_1$ is turned off so that the output photon passes $SM_1$. Figure~\ref{fig:setup}(b) summarizes the on-off states of SM$_1$, SM$_2$, SPR$_1$, and SPR$_2$. It is easy to show that this optical setup is equivalent to Fig.~\ref{fig:convert}, which, compared to Fig.~\ref{fig:scheme}(c), replaces each biased BS by a polarization beam splitter (PBS) together with two polarization rotators (PRs) and each classical switch is replaced by a quantum switch.   

Suppose the quantum switch is initially in the pure state $C_g \ket{g} + C_e \ket{e}$, and hence the overall initial state is $\ket{\psi_0}=\ket{H}(C_g \ket{g} + C_e \ket{e}$).  
After the first PBS, the overall state becomes
\begin{equation}
\ket{\psi_1}=(\cos \beta_1 \ket{H} + \sin \beta_1  \ket{V})(C_g \ket{g} + C_e \ket{e}).
\end{equation}
At the end of the first outer cycle but before the second PBS and its two preceding PRs, the overall state conditioned on the photon not being absorbed by Bob's atom becomes
\begin{equation}
\ket{\psi_1'}=C_g(\cos \beta_1 \ket{H} + \sin \beta_1 \cos^{2N}\beta_2 \ket{V}) \ket{g}+C_e\cos \beta_1 \ket{H}\ket{e} .
\end{equation}
When $N$ goes to infinity, we have $\cos^{2N}\beta_2 = 1$, and hence the overall state can be simplified to $C_g (\cos \beta_1 \ket{H} + \sin \beta_1  \ket{V}) \ket{g}+C_e\cos \beta_1 \ket{H}\ket{e}$.
After the second PBS, the overall state becomes
\begin{eqnarray}
\ket{\psi_2}&=& C_g[\cos (2 \beta_1) \ket{H} + \sin (2 \beta_1) \ket{V}] \ket{g}  \nonumber \\
&& +C_e(\cos^2 \beta_1 \ket{H}+\cos \beta_1\sin \beta_1 \ket{V})   \ket{e} .
\end{eqnarray}
After the $M$th PBS, the overall state becomes 
\begin{eqnarray}
\ket{\psi_M}&=& C_g[\cos (M \beta_1) \ket{H} + \sin (M \beta_1) \ket{V}] \ket{g} \nonumber \\
&&+C_e(\cos^M \beta_1 \ket{H}+\cos^{M-1} \beta_1\sin \beta_1 \ket{V}) \ket{e}. 
\end{eqnarray}
When $M$ goes to infinity, the state can be simplified to $ C_g \ket{V} \ket{g}+C_e\ket{H}\ket{e}$.
Note that since $\ket{g}$ flips the photon state and $\ket{e}$ keeps the photon state unchanged, we view $\ket{g}$ as logical  $\ket{1}$ and $\ket{e}$ as logical $\ket{0}$.
In summary, a counterfactual special CNOT gate performs a quantum transformation
\begin{equation}
\mathcal{P}: (C_g \ket{g} + C_e \ket{e}) \ket{H} \rightarrow C_g \ket{g} \ket{V}+ C_e \ket{e} \ket{H}.
\end{equation}
During this transformation $\mathcal{P}$, there is no interaction between the atom and the photon. The proof is as follows.
First we note that the forward- and backward-evolving wave functions of $\mathcal{P}$ are in a superposition of the ones in Figs.~\ref{fig:twovector}(b) and \ref{fig:twovector}(c). In addition, the forward- and backward-evolving wave functions in Figs.~\ref{fig:twovector}(b) and \ref{fig:twovector}(c)  do not overlap in the transmission channel. Hence, the forward- and backward-evolving wave functions of $\mathcal{P}$ also do not overlap in the transmission channel. This implies that the photon is not present in the channel, and hence there is no interaction between the atom and the photon.

\subsection{Reduction from  a special CNOT gate to a generic CNOT gate}
\label{sec:SingleCompute}

In this section we provide a CUQC scheme based on counterfactual special CNOT gates. Here we restrict our attention to using one atom qubit and multiple photon qubits. In Sec.~\ref{sec:extensions} we will discuss the case of multiple atom qubits.

By Theorem \ref{thm:universal}, single-qubit operations and CNOT gates suffice for universal quantum computation. 
Here the single-qubit operations are local and do not involve interaction among two-level quantum subsystems. A normal CNOT gate,
however, requires two quantum subsystems to interact. One quantum subsystem acts as the control qubit and the other acts as the target qubit. 
The control qubit controls the target qubit. In addition, a normal CNOT gate can be performed on any two qubits and the control qubit can be either of the two qubits. However, a counterfactual special CNOT gate can only be between the atom qubit and the photon qubit, where the atom qubit must be the control qubit.

We show here that by several suitable transformations, any circuits with normal CNOT gates can be transformed to an equivalent circuit with counterfactual special CNOT gates where the control qubits are all atom qubits and the target qubits are all photon qubits.
There are three steps. In the first step, we show that the control qubit and the target qubit can be switched by adding four Hadamard gates. The transformation is illustrated by
\begin{equation}
\centering
\Qcircuit @C=1.5em @R=2em {
 & \targ       & \qw    &  & & \gate{H}& \ctrl{2}        &\gate{H}& \qw \\
 &  &  & =  & & & & &   \\ 
&   \ctrl{-2} &   \qw  &  & &\gate{H} & \targ &\gate{H} & \qw \\
%  & \qw & \ghost{\mathcal{M}(x)}           & \gate{U} & \qw \\
%  & \qw & \ghost{\mathcal{A}(x)}           & \qw & \qw
} 
\label{eq:firsttransform}
\end{equation}

In the second step, we show that if neither the control qubit nor the target qubit is the atom qubit, we can transform the control qubit to the atom qubit as follows (assume the top line is the atom qubit):
\begin{equation}
\centering
\Qcircuit @C=1.5em @R=2em {
\lstick{\ket{0}} &  \qw      & \qw    &  & & \lstick{\ket{0}}& \targ& \ctrl{2}        &\targ& \qw \\
 &  \ctrl{1}  &  \qw & =  & && \ctrl{-1} & \qw & \ctrl{-1}  & \qw   \\ 
&  \targ  &   \qw  &  & & & \qw & \targ &\qw & \qw \\
%  & \qw & \ghost{\mathcal{M}(x)}           & \gate{U} & \qw \\
%  & \qw & \ghost{\mathcal{A}(x)}           & \qw & \qw
} 
\label{eq:secondtransform}
\end{equation}
The first CNOT gate and the third CNOT gate can be further transformed such that the control qubit is switched to the first qubit. 

In the third step, to ensure that the atom qubit always stays at the initial state $\ket{0}$, we can add the atom qubit as an additional qubit to the original circuit. Each time a CNOT gate needs to be performed, the atom qubit goes through the second transformation \eqref{eq:secondtransform} and returns to $\ket{0}$ at the end of the transformation, as shown by
\begin{equation}
\centering
\Qcircuit @C=1.5em @R=2em {
 &      &       & &&\lstick{\ket{0}} &  \qw      & \qw    &  && \lstick{\ket{0}}& \targ& \ctrl{2}        &\targ& \rstick{\ket{0}} \qw\\
 &  \ctrl{1}  &  \qw & \longrightarrow  & &&  \ctrl{1}  &  \qw & \longrightarrow & && \ctrl{-1} & \qw & \ctrl{-1}  & \qw   \\ 
&  \targ  &   \qw  &  & &&  \targ  &   \qw  &  & && \qw & \targ &\qw & \qw \\
} 
\label{eq:thirdtransform}
\end{equation}

By these transformations, the control qubits of all CNOT gates can be concentrated to one qubit. Implementing this qubit  with an atom qubit and all other qubits with photons and changing all CNOT gates to counterfactual special CNOT gates, universal quantum computation can be realized by single-qubit operations and counterfactual special CNOT gates. Since both single-qubit operations and counterfactual special CNOT gates require no interaction among two-level quantum subsystems, a CUQC scheme is achieved.

\section{Examples}
\label{sec:examples}

\begin{figure}[htb]
\centering \includegraphics[width=8.5cm]{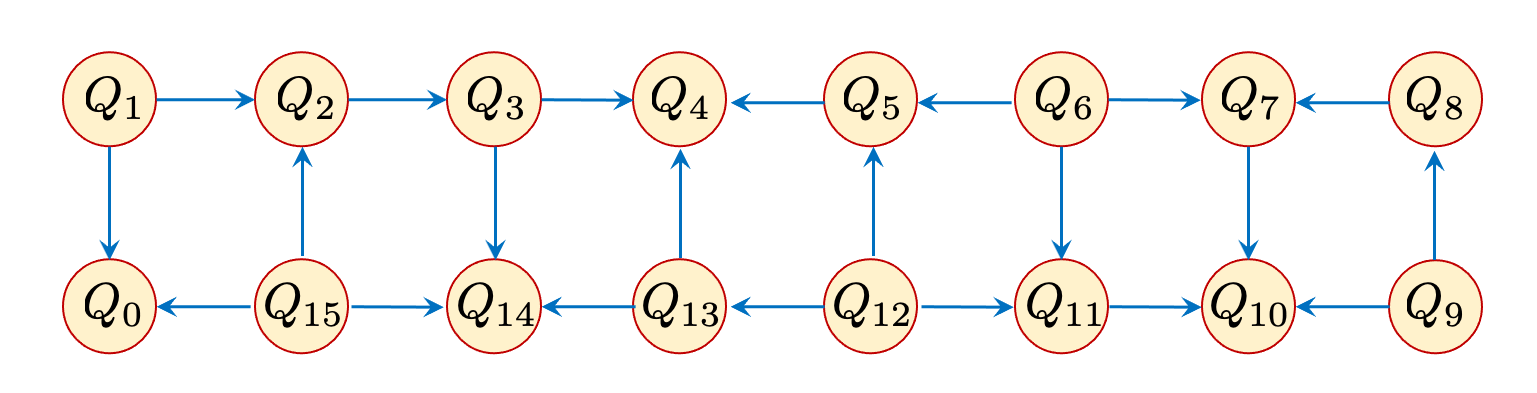}
\caption{Connectivity diagram of {\it ibmqx5}. Here $Q_0, Q_1, \cdots, Q_{15}$ denote the 16 qubits in {\it ibmqx5}.  
An arrow between two qubits shows that these two qubits have control relations. The direction of each arrow is from the control qubit to the target qubit.} 
\label{fig:connectivity}
\end{figure}

In this section we illustrate the power of the CUQC scheme through three examples.
All can be viewed as special cases of our CUQC scheme and may be of independent interest.
 After presenting each example, we will partially demonstrate it on the IBM Q platform \cite{ibmQ}, which contains some small-scale superconducting quantum computers. Note that a full demonstration is left as future work and additionally requires faithful implementation of counterfactual special CNOT gates.  The IBM Q platform provides several processor options, from which we choose {\it ibmqx5} for our demonstration purpose. The option  {\it ibmqx5} features a quantum computer with 16 qubits. The connectivity of these 16 qubits is shown in Fig.~\ref{fig:connectivity}. The symbol $a \rightarrow b$ means that $a$ is the control qubit and $b$ is the target qubit. The quantum circuit given by the user is compiled by the IBM Q compiler so that the circuit can be simulated on IBM's quantum computer.

\subsection{Example 1: Genuine counterfactual communication of a quantum state}
\label{sec:PureQubit}
In the first example, we present the protocol of genuine counterfactual communication of a quantum state.

We start with counterfactually communicating a pure qubit. By applying the counterfactual CNOT gate twice, together with some local operations, the quantum state is transferred from the atom to the photon. The detailed quantum circuit design is shown as follows:
\begin{equation}
\centering
\mathcal{C} =\Qcircuit @C=1.5em @R=2em {
  & \ctrl{1}       & \gate{H}&    \ctrl{1}      &\gate{H}&  \qw &  \\
 &   \targ & \gate{H}&    \targ    &\gate{H}& \qw & 
} 
\label{eq:circuitCommunicate}
\end{equation}

Here, we note that although the goal of counterfactual communication of a quantum state is similar to that of quantum teleportation, in that a quantum state is transferred from one party to another, the two subjects also differ in several important ways. First, in quantum teleportation, it is necessary that the two parties share a Bell state for each qubit to be teleported, while there is no such restriction in counterfactual communication of a quantum state. 
Second, in quantum teleportation, the two parties need to perform classical communication for a successful teleportation, while in our example, such classical communication is not needed.
Therefore, we avoid calling our protocol counterfactual quantum teleportation to emphasize the difference. In the Appendix, we extend our protocol to counterfactual communication of mixed qubits and higher-dimensional quantum states.

We then partially demonstrate the protocol on the IBM Q platform. For simplicity, we test the following
quantum circuit, where we make the assumption that this circuit is compiled into the circuit $\mathcal{C}$ in Eq.~\eqref{eq:circuitCommunicate} and the CNOT gates in this circuit are realized by counterfactual special CNOT gates: 
\begin{equation}
\centering
\Qcircuit @C=1.5em @R=2em {
  & \ctrl{1}      &     \targ   & \qw  \\
 &   \targ&     \ctrl{-1}     & \qw 
} 
\nonumber
\end{equation}
After inputting the states $\ket{00}$ and $\ket{10}$, the returned results of {\it ibmqx5} are $\ket{00}$ and $\ket{01}$. This shows that the first qubit is counterfactually communicated to the second qubit. This example can enable grayscale imaging, by taking Bob's state to be a mixed quantum state.

\subsection{Example 2: Quantum swapping}
In the second example, we consider quantum swapping, which transforms $\ket{x}\ket{y}$ to $\ket{y}\ket{x}$. 
We design the quantum swap gate as follows:
\begin{equation}
\centering
\Qcircuit @C=1.5em @R=2em {
 & \ctrl{1} &  \targ & \ctrl{1}  & \qw  \\
  & \targ& \ctrl{-1}        &\targ& \qw 
%  & \qw & \ghost{\mathcal{M}(x)}           & \gate{U} & \qw \\
%  & \qw & \ghost{\mathcal{A}(x)}           & \qw & \qw
} 
\nonumber
\end{equation}

Recall that by the definition of the quantum swap gate, it keeps $\ket{00}$ and $\ket{11}$ unchanged, and swaps $\ket{01}$ and $\ket{10}$.
After going through the three CNOT gates shown above, the four states become
\begin{eqnarray}
\ket{00} \rightarrow \ket{00} \rightarrow \ket{00} \rightarrow \ket{00},  \nonumber \\
\ket{01} \rightarrow \ket{11} \rightarrow \ket{10} \rightarrow \ket{10},  \nonumber \\
\ket{10} \rightarrow \ket{10} \rightarrow \ket{11} \rightarrow \ket{01},   \\
\ket{11} \rightarrow \ket{01} \rightarrow \ket{01} \rightarrow \ket{11}.  \nonumber 
\end{eqnarray}
Hence the circuit faithfully implements the quantum swap gate.

By switching the control qubit and the target qubit of the second CNOT gate, the circuit can be transformed to
\begin{equation}
\centering
\Qcircuit @C=1.5em @R=2em {
&\ctrl{1}        &\gate{H}&\ctrl{1}&\gate{H}&\ctrl{1}& \qw \\
& \targ & \gate{H}&\targ&\gate{H}&\targ & \qw  
%  & \qw & \ghost{\mathcal{M}(x)}           & \gate{U} & \qw \\
%  & \qw & \ghost{\mathcal{A}(x)}           & \qw & \qw
} 
\nonumber
\end{equation}
By viewing the first qubit as the atom qubit and the second qubit as the photon qubit, such a circuit can be
realized using counterfactual special CNOT gates. This example recovers the result of a recent counterfactual protocol of a quantum swap gate \cite{li2019counterfactual}. 

We then partially demonstrate the above protocol on the IBM Q platform. We again assume that each CNOT gate in the circuit
is realized by a counterfactual CNOT gate.
After inputting the states $\ket{00}$, $\ket{10}, \ket{01}$, and $\ket{11}$, the returned results are $\ket{00}, \ket{01},\ket{10}$, and $\ket{11}$, respectively.
This verifies that such a circuit can counterfactually swap two qubits.

\subsection{Example 3: Quantum error erasure code}
In the third example, we consider quantum error erasure codes.
A quantum error erasure code is a type of error correction code that focuses on recovering erasure errors. 
The smallest quantum error erasure code contains four qubits.
The encodings of logic 0 and logic 1 are 
\begin{eqnarray}
\ket{0_L} & =  &\ket{0000} + \ket{1111},  \nonumber  \\
\ket{1_L}  & =  & \ket{1001} + \ket{0110}.
\end{eqnarray}
To prove that this code can correct erasure errors, it suffices to prove that it can correct bit flip errors in two bases that can be transformed by Hadamard gates \cite{steane1996error}. 
The base that is transformed from the encoding basis with Hadamard gates is
\begin{eqnarray}
\centering
\ket{0^\bot} & =  &\ket{0000} + \ket{0011}  + \ket{0101} + \ket{0110} \nonumber  \\
& & +\ket{1001} + \ket{1010}  + \ket{1100} + \ket{1111}, \nonumber  \\
\ket{1^\bot} & =  &\ket{0000} - \ket{0011}  - \ket{0101} + \ket{0110} \\
& & +\ket{1001} - \ket{1010}  - \ket{1100} + \ket{1111}. \nonumber
\end{eqnarray}

In the error erasure channel, the position of the error is known but the correct value is unknown.
Note that the parity is even in the logic encoding; hence an odd encoding indicates an error and 
this error can be corrected by properly choosing the bit value on the erroneous qubit so that the parity is restored.
In terms of quantum circuit design, one can design the following circuit to encrypt $\ket{\psi} = \alpha \ket{0}  + \beta \ket{1}$ to
$\alpha \ket{0_L}  + \beta \ket{1_L}$:
\begin{equation}
\centering
\Qcircuit @C=1.5em @R=2em {
\lstick{\ket{\psi}} &  \ctrl{3}&\qw       &\targ& \qw & \qw\\
\lstick{\ket{0}}  & \qw & \gate{H} & \ctrl{-1} \qwx[1]  &\ctrl{2} & \qw  \\ 
\lstick{\ket{0}}  &\qw & \qw &\targ & \qw& \qw \\
\lstick{\ket{0}}  & \targ& \qw &\qw &\targ & \qw 
} 
\nonumber
\end{equation}

After the first gate, $\ket{0}$ becomes $\ket{0000}$ and $\ket{1}$ becomes $\ket{1001}$. After the second gate,
$\ket{0000}$ becomes $\ket{0} \otimes (\ket{0}+\ket{1}) \otimes \ket{00}$ and $\ket{1001}$ becomes $\ket{1} \otimes (\ket{0}+\ket{1}) \otimes \ket{01}$.  
After the third, fourth and fifth gates, $\ket{0} \otimes (\ket{0}+\ket{1}) \otimes \ket{00}$ becomes $\ket{0000} + \ket{1111}$ and $\ket{1} \otimes (\ket{0}+\ket{1}) \otimes \ket{01}$ becomes $\ket{1001} + \ket{0110}$. This shows that the encoding by the circuit achieves the logic encoding. 

By the second transformation \eqref{eq:secondtransform} and taking the second qubit to be the atom qubit, the circuit is turned into 
\begin{equation}
\centering
\Qcircuit @C=1.5em @R=2em {
\lstick{\ket{\psi}} &  \ctrl{1}& \qw&  \ctrl{1}&\qw       &\targ& \qw & \qw\\
\lstick{\ket{0}}  &\targ& \ctrl{2}& \targ& \gate{H} & \ctrl{-1} \qwx[1]  &\ctrl{2} & \qw  \\ 
\lstick{\ket{0}}  &\qw&\qw&\qw & \qw &\targ & \qw& \qw \\
\lstick{\ket{0}}  & \qw & \targ & \qw& \qw &\qw &\targ & \qw 
} 
\nonumber
\end{equation}
The first and third CNOT gates can be further transformed using the first transformation \eqref{eq:firsttransform}, after which all the control qubits are set to the second qubit. Hence, the logic encoding of a quantum erasure code can be counterfactually prepared.

We then partially demonstrate this example on the IBM Q platform. We make the same assumption that the CNOT gates are realized by
counterfactual special CNOT gates. When inputting the states $\ket{0000}$ and $\ket{1000}$, the returned results are $\ket{0000}+\ket{1111} and \ket{0110}+\ket{1001}$. This verifies that the encodings of the logical 0 and 1 are counterfactually prepared.

\section{Practical issues}
\label{sec:extensions}
In this section we consider some practical aspects of the CUQC scheme.
In particular, we give some variants of CUQC that address practical issues such as efficiency improvement and device imperfections. We first replace the single atom in the standard CUQC by multiple atoms and show to what
extent such a modification improves the overall system efficiency. Next we show the effect of device imperfections 
on the performance of the CUQC scheme. The device imperfections 
include finite $M$, finite $N$, photon loss, and  atom missing.
 
\subsection{CUQC with multiple atoms}
\label{sec:multi}
In this section we show that multiple atoms can  significantly reduce the circuit depth in some cases. 

Suppose a circuit contains $n$ qubits and these $n$ qubits are grouped into $n/2$ pairs of qubits. For each pair of qubits, there is a CNOT gate from one of the qubits to the other qubit in the circuit.
In total, there are $n/2$ pairs of parallel CNOT gates in this circuit and there are no other gates. 

Now, we consider the implementation of this circuit. In the case that there is only one atom qubit,
by the second transformation \eqref{eq:secondtransform}, this atom qubit needs to process $3n/2$ CNOT gates and hence the circuit depth is $\Omega(n)$.
In contrast, if there are $n/2$ atom qubits, the circuit depth can be significantly reduced to at most $3$ (each atom qubit handles one CNOT gate). 

Note that the total gate number of the circuit remains similar regardless of the number of atoms. This type of speedup is in the same vein as parallel computing. The atom qubits here take the place of central processing units in parallel computing architecture.

\subsection{Error analysis}
\label{sec:error}

In previous sections, we have assumed ideal devices and infinite cycles. 
In this section, we consider various practical aspects, including finite $M$, finite $N$, photon loss,  and atom missing.

We consider two figures of merits: the efficiency of the computation and the fidelity of the computed result. Since these two figures  
depend strongly on the number of counterfactual special CNOT gates in the circuit, we first analyze the case where the circuit consists of a single
counterfactual special CNOT gate. After that, we discuss quantum circuits that contain multiple counterfactual special CNOT gates.

For the first part, we consider a single counterfactual special CNOT gate in the case of practical devices.
Recall that in our setup, the outer cycle rotates the polarization by $\beta_1 = \pi /2M$ and the inner cycle rotates the polarization by 
$\beta_2 = \pi /2N$. There are $M-1$ outer cycles and for each outer cycle there are $2(N-1)$ inner cycles. 

Let $P_D$ denote the probability that the photon passes the transmission channel. The efficiency is defined as $\textsf{E}=1-P_D$. 
The unnormalized output state of the circuit is defined by 
\begin{equation}
\ket{\psi_{\footnotesize\textrm{final}}}=C_1 \ket{H}\ket{g} + C_2 \ket{V}\ket{g} + C_3\ket{H}\ket{e} + C_4\ket{V}\ket{e}.
\end{equation}
The efficiency $\textsf{E}$ can be represented as 
 \begin{equation}
\textsf{E}=\sum_{i=1}^4 |C_i|^2.
\label{eq:efficiency}
\end{equation}
 Since the ideal output is $\ket{\psi_{\footnotesize\textrm{ideal}}}=C_e\ket{H}\ket{e} + C_g \ket{V}\ket{g}$, the fidelity is
\begin{equation}
\textsf{F}=| C_e^* C_3 + C_g^* C_2|^2 /\textsf{E}.
\label{eq:fidelity}
\end{equation}

We proceed by first analyzing the final state of the photon when the atom is in the state $\ket{e}$ under finite $M$ and 
$N$. In this case, the initial photon state $\ket{H}$ will become $\cos^M \beta_1 \ket{H}$ at the end. Now we analyze the
more complicated case that the atom is in the state $\ket{g}$.
Let $(X, Y)^T$ denote the quantum state $X\ket{H}+Y\ket{V}$.
For each outer cycle, the state of the photon evolves
as 
\begin{equation}
\left(    
 \begin{array}{c}  
    X_i  \\ 
    Y_i \\
  \end{array}
\right)
=
\left(
\begin{array}{cc}  
    1 & 0\\ 
   0 & \cos^{2N} \beta_2\\
  \end{array}  
  \right)
  \left(
  \begin{array}{cc}  
    \cos \beta_1 & \sin\beta_1\\ 
   -\sin\beta_1& \cos \beta_1\\
  \end{array}  
  \right)
\left(  
 \begin{array}{c}  
    X_{i-1} \\ 
    Y_{i-1} \\
  \end{array}
  \right)
\end{equation}
The initial conditions are $X_0 = 1$ and $Y_0 = 0$ since the initial photon state is $\ket{H}$.
Hence, for an initial state 
\begin{equation}
\ket{\psi_{\footnotesize\textrm{initial}}}=\ket{H}(C_e\ket{e} + C_g\ket{g}),
\end{equation}
 the final state becomes
\begin{equation}
\ket{\psi_{\footnotesize\textrm{final}}}=C_e\cos^M \beta_1  \ket{H}\ket{e} + C_g X_M\ket{H}\ket{g}+C_g Y_M\ket{V}\ket{g}.
\end{equation}
The efficiency and the fidelity are then calculated according to Eqs.  \eqref{eq:efficiency} and \eqref{eq:fidelity} .

\begin{figure}[htb]
\centering \includegraphics[width=8.5cm]{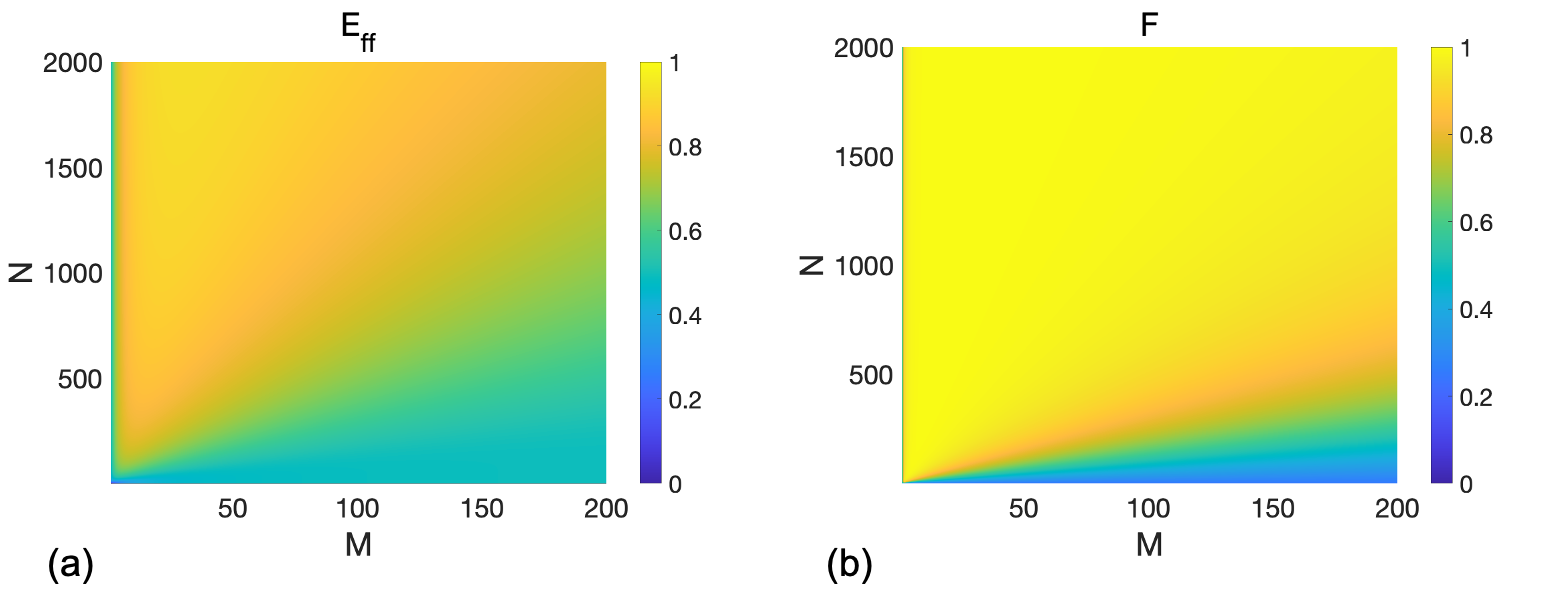}
\caption{(a) Efficiency of the circuit under various $M$ and $N$. The efficiency is almost the same for the same ratio of $N$ to $M$ and  becomes higher when the ratio of $N$ to $M$ increases. However, for a $N/M$ ratio of 20, the efficiency is still low (below 0.9). (b) Fidelity of the circuit output under various $M$ and $N$. The fidelity is approximately the same for the same ratio of $N$ to $M$ and 
becomes higher when the ratio of $N$ to $M$ increases. When the $N/M$ ratio is above 5, the fidelity is close to unity.} 
\label{fig:eff}
\end{figure}

For an initial atom state $(\ket{g}+\ket{e})/\sqrt{2}$, we plot the efficiency against different $M$ and $N$ in Fig.~\ref{fig:eff}(a) and the fidelity against different $M$ and $N$ in Fig.~\ref{fig:eff}(b). It can be seen 
that the fidelity approaches one when $N$ is over five times $M$. The efficiency is more stringent
for the ratio of $N$ to $M$. Even when the ratio of $N$ to $M$ is 20, the efficiency is still below 0.9.

Hereafter, we set $M=10$ and $N=200$ and consider the effect of photon loss and atom missing. We let $\gamma$ denote the probability of photon loss in the transmission channel  and let $\eta$ denote the probability of atom missing. 

\begin{figure}[htb]
\centering \includegraphics[width=8.5cm]{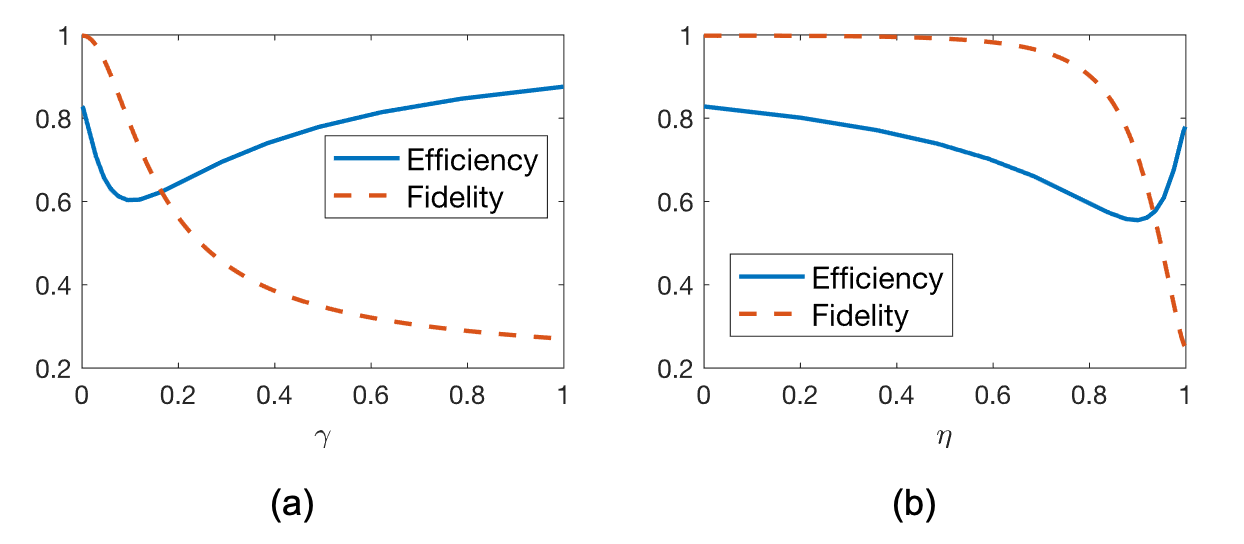}
\caption{ (a) Efficiency (blue solid line) and fidelity (red dashed line) vs the photon loss probability $\gamma$. When $\gamma$ increases in the range $0<\gamma<0.1$, the efficiency and the fidelity decrease
rapidly. This shows that the efficiency and the fidelity are quite sensitive to a small probability of photon loss. (b) Efficiency and fidelity vs the atom missing probability $\eta$. When $\eta$ increases in the region $0<\eta<0.1$, the efficiency
and the fidelity degrade slowly. This shows that the protocol is resilient to infrequent block failure of the atom.} 
\label{fig:miss}
\end{figure}

We first analyze the effect of photon loss and assume there is no atom missing. If the atom is in the ground state, the transmission is always blocked; hence whether the photon is lost
in the transmission channel does not make any difference. Therefore, we only need to consider the 
case that the atom is in the state $\ket{e}$. After $N-1$ inner cycles, some proportion of the vertical component will go into the next $N-1$ inner cycles. The proportion is given by $W$, the first entry of the vector
\begin{equation}
\left(  
 \begin{array}{c}  
   W \\ 
   * \\
  \end{array}
  \right)
  =
\left(
  \begin{array}{cc}  
    \cos \beta_2 & \sin\beta_2\\ 
   -(1-\gamma)\sin\beta_2& (1-\gamma)\cos \beta_2\\
  \end{array}  
  \right)^{N}
\left(  
 \begin{array}{c}  
   1 \\ 
    0 \\
  \end{array}
  \right).
\end{equation}
After being reflected by the double-sided mirror, this vertical component goes through another $N-1$ inner cycles before 
reaching the end of one outer cycle,
and hence the $W^2$ proportion of the vertical component remains after each outer cycle.
The state change for each outer cycle becomes
\begin{eqnarray}
 \left(    
 \begin{array}{c}  
    U_i  \\ 
    V_i \\
  \end{array}
\right) 
=
\left(
\begin{array}{cc}  
    1 & 0\\ 
   0 & W^2\\
  \end{array}  
  \right) 
\left(
  \begin{array}{cc}  
    \cos \beta_1 & \sin\beta_1\\ 
   -\sin\beta_1& \cos \beta_1\\
  \end{array}  
  \right)
\left(  
 \begin{array}{c}  
    U_{i-1} \\ 
    V_{i-1} \\
  \end{array}
  \right). 
\end{eqnarray}
The initial conditions are $U_0 = 1$ and $V_0 =0$.
After $M-1$ outer cycles, the final state becomes 
\begin{equation}
\ket{\psi_{\footnotesize\textrm{final}}}=C_e U_M   \ket{H}\ket{e} +C_e V_M   \ket{V}\ket{e} + C_g X_M\ket{H}\ket{g}+C_g Y_M\ket{V}\ket{g}.
\end{equation}
We plot the efficiency and fidelity as a function of the 
photon loss probability $\gamma$ in Fig.~\ref{fig:miss}(a). It can be seen that the fidelity steadily
decreases, while the efficiency initially declines and then slowly increases. When the photon loss is small, both the fidelity and the efficiency decrease significantly as the photon loss increases.

Now we analyze the effect of atom missing and assume that there is no photon loss. In the atom missing case, the atom may fail to block the photon and let the photon pass through. Apparently, 
if the atom is in the state $\ket{e}$, whether the atom blocks does not make any difference.
When the atom is in the state $\ket{g}$, the vertical component no longer decays with a decay rate $\cos^{2N} \beta_2$ for each outer cycle.

After $N-1$ inner cycles, the vertical component decays as the first entry $Z$ in the vector
\begin{equation}
\left(  
 \begin{array}{c}  
   Z \\ 
   * \\
  \end{array}
  \right)
  =
\left(
  \begin{array}{cc}  
    \cos \beta_2 & \sin\beta_2\\ 
   -\eta\sin\beta_2& \eta\cos \beta_2\\
  \end{array}  
  \right)^{N}
\left(  
 \begin{array}{c}  
   1 \\ 
    0 \\
  \end{array}
  \right).
\end{equation}
After being reflected by the double-sided mirror and going through another $N-1$ inner cycles, the vertical component decays with the same proportion,
so the overall decay for one outer cycle is $Z^2$.

When $\eta=0$, this first entry is $Z=\cos^{N} \beta_2$, recovering the ideal case.
The state change for each outer cycle becomes
\begin{eqnarray}
\left(    
 \begin{array}{c}  
    X_i'  \\ 
    Y_i' \\
  \end{array}
\right) 
=
\left(
\begin{array}{cc}  
    1 & 0\\ 
   0 &Z^2\\
  \end{array}  
  \right) 
\left(
  \begin{array}{cc}  
    \cos \beta_1 & \sin\beta_1\\ 
   -\sin\beta_1& \cos \beta_1\\
  \end{array}  
  \right)
\left(  
 \begin{array}{c}  
    X_{i-1}' \\ 
   Y_{i-1}' \\
  \end{array}
  \right). 
\end{eqnarray}
The initial conditions are $X_0 '= 1$ and $Y_0' =0$.
After $M-1$ outer cycles, the final state becomes 
\begin{equation}
\ket{\psi_{\footnotesize\textrm{final}}}=C_e\cos^M \beta_1  \ket{H}\ket{e} + C_g X_M'\ket{H}\ket{g}+C_g Y_M'\ket{V}\ket{g}.
\end{equation}
We plot the efficiency and the fidelity as a function of the atom missing probability $\eta$ in Fig.~\ref{fig:miss}(b). It can
be seen that the efficiency and the fidelity both decrease slowly as $\eta$ increases in the range $0<\eta <0.1$.

% Key question 2: how the composition of the CNOT gate affects the error accumulation.
For the second part, we show how the composition of the CNOT gates affects the transfer efficiency
and the fidelity of a circuit with multiple counterfactual special CNOT gates.
Suppose there are $K$ CNOT gates in a circuit $\textsc{Circ}$. The transfer efficiency
and the fidelity for a single CNOT gate are denoted by $\textsf{E}(\textrm{CNOT})$ and $\textsf{F}(\textrm{CNOT})$, respectively.

Since \textsc{Circ}  gets an output if and only if all CNOT
gates in \textsc{Circ} succeed, the efficiency of $\textsc{Circ}$ is 
\begin{equation}
\textsf{E}(\textsc{Circ})=\textsf{E}(\textrm{CNOT})^K.
\end{equation}
For the fidelity part, we can view each quantum state as a vector and 
the deviation can be characterized by the angle $\theta$ between
the ideal vector and the actual vector. By the definition of fidelity,
we have 
\begin{equation}
\textsf{F}(\textrm{CNOT}) = \cos^2 \theta.
\label{eq:fidelityangle}
\end{equation}
 In the worst case, the deviation happens
in the same direction for all $K$ CNOT gates. In that case, the final
vector deviates from the original vector by an angle of $K\theta$. 
Combined with Eq.~\eqref{eq:fidelityangle}, the fidelity of $\textsc{Circ}$ is lower bounded
by 
\begin{equation}
\textsf{F}(\textsc{Circ}) \ge \cos^2[K \arccos \sqrt{\textsf{F}(\textrm{CNOT})}].
\end{equation}

% The composition of transfer efficiency is a product.

% The composition of fidelity is also a product.  

\section{Conclusion}
\label{sec:conclusion}
In summary, we have shown a CUQC scheme of which the two-level quantum subsystems have no interaction with each other during the computation. The CUQC scheme is based on counterfactual special CNOT gates together with appropriate circuit transformations. On a physical level, the scheme is based on an atom qubit and an arbitrary number of photon qubits. As special cases, we have shown counterfactual communication of a quantum state, counterfactual quantum state swapping, and counterfactual quantum erasure code. The CUQC scheme has been extended to the multiple atom qubit case,  the case of a finite round number, and the case of imperfect devices such as photon loss and atom missing. 

The CUQC scheme illustrates again the mysterious nature of quantum physics.
In addition to its theoretical interest, this scheme also has some practical applications. In particular, the scheme enables grayscale imaging, by taking the atom state in Sec.~\ref{sec:PureQubit} to be $p_i \ket{0}\bra{0}+(1-p_i)\ket{1}\bra{1}$, where $0\le p_i\le 1$ characterizes the grayness of the image. By combining the grayscale images  of three primary colors, we recover the full color image. The imaging method we propose here has the notable feature that no photon touches the image. This is especially important for the imaging of ancient arts, to which even faint light can potentially cause significant damage. 

As future directions, it would be interesting to explore optical elements that can split into more than two paths. In that case, a qutrit or a qudit, for example, can be counterfactually transmitted with the same number of rounds as a qubit.  Another direction is to design  passive substitutes for the switchable components of a counterfactual special CNOT gate, as this part is quite difficult to realize in laboratories. A third direction is to find other physical platforms to realize the CUQC scheme, in addition to the hybrid system of atoms and photons. For example, one can consider whether an all-optical system can realize the CUQC scheme. As a fourth direction, a full experimental realization of the CUQC scheme and its special cases presented in this work is of great interest.

\section*{Acknowledgements}
This work was supported by the internal Grant No. SLH00202007 from East China University of Science and Technology.

\appendix

\section{Genuine counterfactual communication of mixed qubits and higher dimensional quantum states}
\label{Appsec:Mixed} 
In this appendix, we generalize our counterfactual communication result first to transmitting a mixed qubit and then to transmitting higher-dimensional quantum states.

Note that a mixed qubit can be represented by $\{ (p_1, \ket{\psi_{1A}}), (p_2, \ket{\psi_{2A}}), \cdots, (p_k, \ket{\psi_{kA}}) \}$, where $p_i $ is the probability of the state  $\ket{\psi_{iA}} $ and the subscript $A$ means that this is an atom qubit. 
During the protocol, the whole system evolves as
\begin{eqnarray}
 && \{ (p_1, \ket{\psi_{1A}}), \cdots, (p_k, \ket{\psi_{kA}}) \} \ket{H}  \nonumber  \\
&= & \{ (p_1, \ket{\psi_{1A}} \ket{H}),  \cdots, (p_k, \ket{\psi_{kA}} \ket{H}) \} \nonumber \\
&\stackrel{\mathcal{P}}{\rightarrow}  & \{ (p_1, \ket{e}\ket{\psi_{1P}} ), \cdots, (p_k, \ket{e}\ket{\psi_{kP}} ) \}   \nonumber \\
&= &  \ket{e}\{ (p_1,\ket{\psi_{1P}} ), \cdots, (p_k,\ket{\psi_{kP}} ) \},  
\end{eqnarray}
where $\ket{\psi_{iA}}=\ket{\psi_{iP}}$, $\mathcal{P}$ is the action of the protocol, and the subscript $P$ in $\ket{\psi_{iP}}$ means that this is a photon qubit. 
Hence, by applying the quantum protocol, the state of the photon becomes $\{ (p_1, \ket{\psi_{1P}}), (p_2, \ket{\psi_{2P}}), \cdots, (p_k, \ket{\psi_{kP}})\}$. In other words, the mixed quantum state from the atom has transmitted to the photon faithfully.

Next we generalize the qubit result to higher dimensions. 
We note that higher-dimensional quantum states can be viewed as multiple qubit states that are entangled. For simplicity, 
 we show that two entangled qubits can be counterfactually  communicated. The case of more than two qubits is similar.
 Two entangled qubits can be represented in the computational basis as
 \begin{equation}
C_{ee} \ket{ee} +C_{eg}   \ket{eg}  +  C_{ge} \ket{ge} +C_{gg} \ket{gg}.
 \end{equation}
 During the protocol, the whole system evolves as
 \begin{eqnarray}
 &&( C_{ee} \ket{ee} + C_{eg}   \ket{eg}  +  C_{ge}  \ket{ge}+ C_{gg} \ket{gg} ) \ket{HH} \\
&= & C_{ee} \ket{ee} \ket{HH} + C_{eg}   \ket{eg} \ket{HH} +  C_{ge}  \ket{ge}  \ket{HH}+ C_{gg} \ket{gg} \ket{HH}\nonumber  \\
&\stackrel{\mathcal{P}}{\rightarrow}  & C_{ee} \ket{ee} \ket{HH} + C_{eg}   \ket{ee} \ket{HV}  + C_{ge}  \ket{ee}  \ket{VH}+ C_{gg} \ket{ee} \ket{VV}  \nonumber \\
&= &  \ket{ee} ( C_{ee} \ket{HH} + C_{eg}   \ket{HV}+  C_{ge}  \ket{VH}+ C_{gg} \ket{VV} ),  \nonumber 
\end{eqnarray}
where $\mathcal{P}$ is the action of the protocol.
In other words, after the protocol, the state of the atoms is transferred to the photons faithfully and turns the state of the photons to $ C_{ee} \ket{HH} + C_{eg}   \ket{HV}+  C_{ge}  \ket{VH}+ C_{gg} \ket{VV} $. 
 Similar to the previous analysis that extends pure qubits to mixed qubits, the analysis of high-dimensional pure quantum states can also be extended to that of high-dimensional mixed  quantum states.

%%%%%%%%%%%%%%%%%%%%%%%%%%%%%%%%%%%%%%%%
% choose a style
%\bibliographystyle{ieeetr}
%\bibliographystyle{unsrt}
\bibliographystyle{apsrev4-1}
% \bibliographystyle{iopart-num}
%%%%%%%%%%%%%%%%%%%%%%%%%%%%%%%%%%%%%%%%

%%%%%%%%%%%%%%%%%%%%%%%%%%%%%%%%%%%%%%%%
% choose a .bib file
\bibliography{BibliWeak}
%%%%%%%%%%%%%%%%%%%%%%%%%%%%%%%%%%%%%%%%

\end{document}